\documentclass{aastex}
\usepackage{emulateapj5}
\usepackage[latin1]{inputenc}
\usepackage{amsmath,amssymb,graphicx,natbib}
\def\kms{\ifmmode{{\rm km}\,{\rm s}^{-1}}\else{km\,s$^-1$}\fi}
\def\Mpc{{\rm Mpc}}
\def\rb{{\rm b}}
\def\rk{{\rm k}}
\def\zr{z_{\text{re}}}

\begin{document}

\title{Constraints on the redshift of reionization from CMB data}

\author{
Jens Schmalzing\altaffilmark{1},
Jesper Sommer-Larsen, and
Martin Götz
}
\affil{
Teoretisk Astrofysik Center,
Juliane Maries Vej 30,
DK-2100 K\o benhavn \O,
Denmark.
}
\affil{
jens,jslarsen,gotz@tac.dk
}
\altaffiltext{1}{
{\em on leave from:}
Ludwig-Maximilians-Universität,
Theresienstraße 37,
D-80333 München,
Germany.
}

\begin{abstract}
We use the recent CMB power spectrum measurement by the Maxima
experiment {\citep{hanany2000}} to constrain the redshift of
reionization $\zr$.  This becomes possible by combining the CMB data
with cosmological parameters from various independent measurements,
including Big Bang Nucleosynthesis (BBN) and X-ray cluster data.  Most
notably, our results provide a robust lower bound on $\zr$.  We find
that $\zr>15$ (8) at the 68\% (95\%) confidence level, unless the
Hubble constant is larger than 75 $\kms\Mpc^{-1}$.
\end{abstract}


\section{Introduction}
\label{sec:introduction}

In CDM cosmologies the first baryonic objects form at redshifts as
high as 50 (see, e.g., {\citealt{haiman1996}} and references
therein). At some lower redshift $\zr$ the formation of these and
subsequent baryonic objects leads to the reionization of the universe
by the combined effects of the fairly hard UV radiation from AGNs and
the softer UV radiation from young, star-bursting galaxies and
possibly also from a population of early, more uniformly distributed
stars (the putative population III stars).  Typical estimates of $\zr$
for CDM models lie in the range $\zr\sim10-20$ {\citep{haiman1998}}.

Recently, various problems with CDM on galactic scales have lead to
the proposal that dark matter is ``warm'' rather than ``cold''
(\citealt{moore1999,hogan1999,sommer-larsen1999} and references
therein). In WDM models the redshift of formation of the first
baryonic objects is smaller than for CDM and increasingly so with
decreasing WDM particle mass. It obviously still provides an upper
limit to the redshift of reionization. Observational lower limits on
the redshift of reionization can therefore be used to place lower
limits to the mass of the putative warm dark matter particles.

Spectral observations of quasars at $z\la5$ show evidence for a
relatively early epoch of reionization. The lack of complete
absorption shortward of the Ly$\alpha$ line (the so-called
Gunn-Peterson effect; {\citealt{gunn1965}}) implies $\zr\ga4-5$,
possibly even $\zr>5.8$ {\citep{fan2000}}.

Constraints on $\zr$ can also be obtained from data on anisotropies of
the cosmic microwave background (CMB) radiation
{\citep{griffiths1999}}. Compton scattering of CMB photons on free
electrons leads to the suppression of the primary anisotropies on
small and intermediate angular scales and hence a characteristic
change of the angular power-spectrum $C_\ell$.

So far the CMB data have been used to place upper limits on the
redshift of reionization $\zr\la30$ {\citep{griffiths1999}}. In this
{\em Letter} we use the Maxima-1 and combined Boomerang and Maxima-1
data {\citep{hanany2000,jaffe2000}} to place astrophysically
interesting upper {\em and} lower limits on $\zr$. The {\em Letter} is
organized as follows: In Section~\ref{sec:parameters}, we motivate the
choice of a number of cosmological parameters prior to analyzing the
CMB data.  Section~\ref{sec:method} briefly explains our method of
analysis, the results of which are summarized in
Section~\ref{sec:results}.  Section~\ref{sec:discussion} discusses
those results and provides an outlook.

\section{Cosmological parameters}
\label{sec:parameters}

Apart from the redshift of reionization that we are interested in, a
cosmological model is specified by a number of other parameters.
Unfortunately, today's CMB measurements alone still provide rather
broad constraints on them when they are jointly taken into account (an
extreme example of the astonishing consequences can be found in Figure
1 of {\citealt{tegmark2000:towards}}).  To obtain meaningful results,
one has to use other observational data, or one's own prejudice, to
fix some of the parameters beforehand.

To begin with, we have to specify what sort of energy is present in
the Universe.  Usually, one takes into account five different kinds,
namely vacuum energy $\Omega_\Lambda$, space curvature $\Omega_\rk$,
baryonic matter $\Omega_\rb$, hot dark matter $\Omega_{\text{HDM}}$,
and cold dark matter $\Omega_{\text{CDM}}$, where the values denoted
by $\Omega$ are the energy densities of the respective type relative
to the critical density.  On the basis of standard inflationary theory
we assume that there is no space curvature, $\Omega_\rk=0$, and also
that the initial power spectrum is scale invariant and no tensor modes
are present (see, however, the discussion in
Section~\ref{sec:discussion}).  Furthermore, dark matter shall be cold
or warm, $\Omega_{\text{HDM}}=0$, since the effect of massive
neutrinos will be negligible on the angular power spectrum of the CMB
{\citep{dodelson1996}} -- but see below.  Obviously, this means that
only three types of energy remain in the Universe.  They are linked
via the relation $\Omega_\Lambda+\Omega_{\text{CDM}}+\Omega_\rb=1$.

The density in baryons $\Omega_\rb$ can be constrained rather nicely
using Big Bang Nucleosynthesis (BBN).  Taking into account a broad
range of recent measurements {\citep{tytler2000}}, it seems reasonable
to set the value of $\Omega_{\rb}h^2$ to $0.019\pm0.002$, where $h$
denotes the Hubble constant measured in units of 100 $\kms\Mpc^{-1}$.

Furthermore, measurements on bright X-ray clusters
{\citep{ettori1999}} provide a robust estimate of the baryonic
fraction $f_\rb$, the ratio of gas to total matter in clusters.  Both
observations {\citep{bahcall1995:where}} and simulations
{\citep{white1993:baryon,evrard2000}} suggest that this is more or
less equal to the fraction of baryons versus all matter in the whole
Universe.  Translated into our cosmological parameters, we have
\begin{equation}
f_\rb
=
\frac{\Omega_\rb}{\Omega_\rb+\Omega_{\text{CDM}}} 
=
(0.069\pm0.012)h^{-3/2}+0.04.
\label{eq:baryonnicfraction}
\end{equation}

Finally, the Hubble constant is taken to be $h=0.65\pm0.10$, as a
plausible compromise between various recent estimates
(\citealt{perlmutter1999,mould2000,reese2000,patel2000,freedman2000}
and references therein).

The possible contribution to the dark matter density from massive
neutrinos with free-streaming mass scale much larger than the mass of
large clusters of galaxies would not be accounted for by this method.
Because of the small mass of such neutrinos ($\la1\text{eV}$), the
contribution to $\Omega_{\text{DM}}$ can be neglected.  That
$\Omega_{\text{HDM}}\ll1$ is also a possible interpretation of the
Super-Kamiokande atmospheric neutrino data, as discussed by
{\citet{tegmark2000:towards}}.

\section{Method}
\label{sec:method}

Recently, {\citet{hanany2000}} published a measurement of the power
spectrum of the CMB from a highly resolved patch of the sky.  We use
their published data throughout.  We have also used the combined
Boomerang and Maxima data of {\citet{jaffe2000}} for our analysis and
have obtained results very similar to the ones presented here.

The measured CMB data consists of ten estimates of the power spectrum
$C_\ell$ averaged over a certain $\ell$ range.  In order to compare
the measured CMB data to a given cosmological model, we first
calculate a numerical spectrum with CMBFAST
{\citep{seljak1996:lineofsight}}.  The model spectrum is then binned
in the same way as the observed data, and the resulting averages
compared to the data by calculating $\chi^2$.  For measurements
following Gaussian statistics, we have
\begin{equation}
\chi^2
=
\sum_{i=1}^{10}\left(
\frac{\Delta^{\text{(data)}}_i-\Delta^{\text{(model)}}_i}{\sigma_i}
\right)^2,
\end{equation}
where $\Delta^{\text{(data)}}_i$ and $\Delta^{\text{(model)}}_i$
denote the average power in the $i$th bin for the Maxima data and the
given model, respectively, and $\sigma_i$ is the measurement error of
the data point.  Actually, the probability distribution function of
power spectrum measurements is slightly skewed in comparison to a
Gaussian, so we correct for this effect with the offset lognormal
method by {\citet{knox1998,bond2000}}.

Assuming that the individual power measurements are independent, a
confidence level of 68\% corresponds to a $\chi^2$ of less than 11.5,
while a $\chi^2$ of 18.3 or more allows to reject a model at the 95\%
confidence level.  Therefore, we always show the contours of these
$\chi^2$ values below.  To put it simply, we ask the question ``Which
models fit well?''

Normally, however, one does not perform this simple $\chi^2$ analysis.
Instead, errors are assigned to fitted parameters by maximum
likelihood.  Most importantly, this means that the error is determined
in comparison to the best fit, and will in general be smaller.  Again,
one can summarize this approach in a simple question: ``Which model
fits best?''

\section{Results}
\label{sec:results}

\begin{figure}
\includegraphics[width=\linewidth]{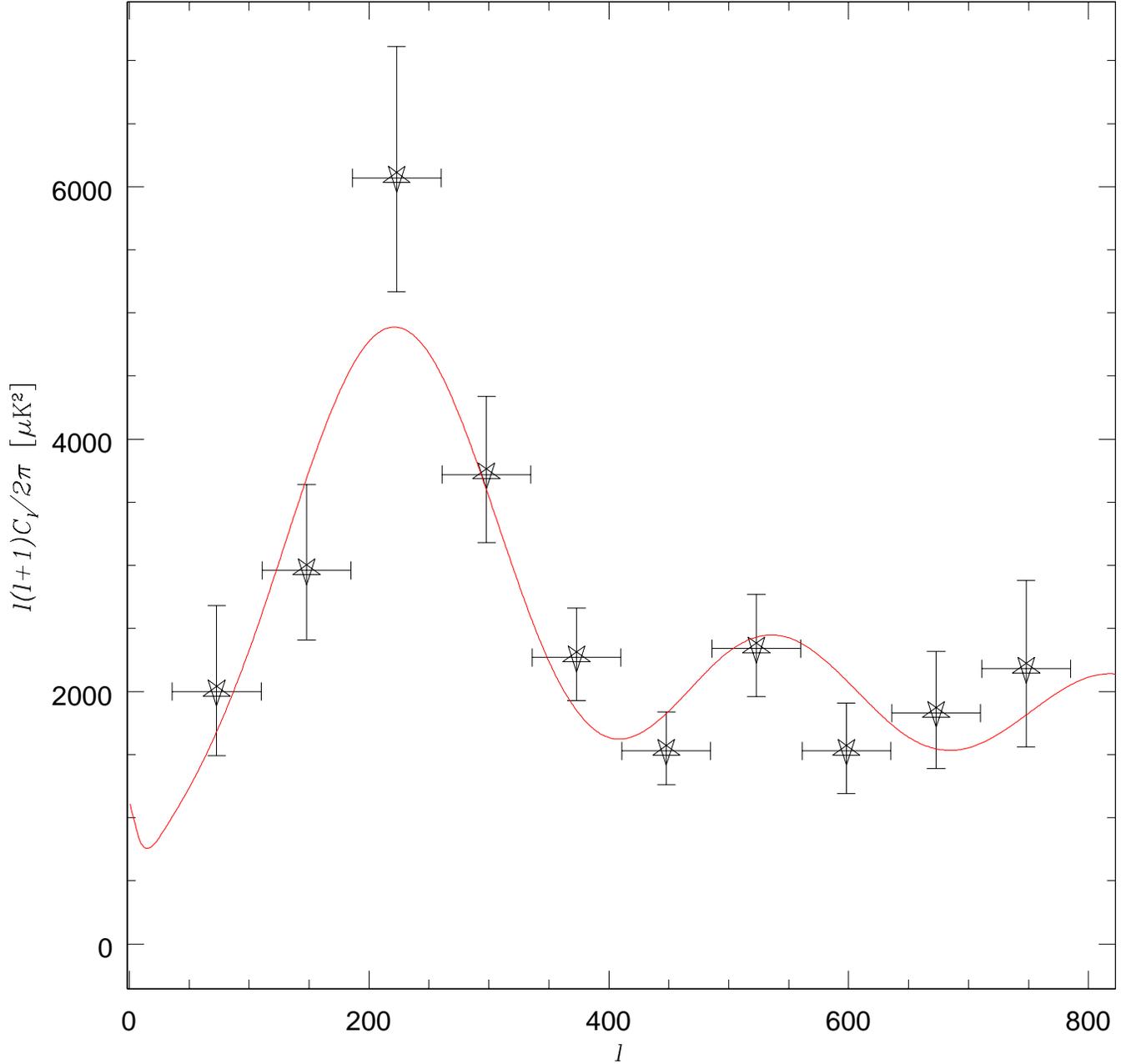}
\caption{
\label{fig:maxima_bestfit}
The best fit (red line) to the Maxima data (black crosses).  The
length of the crosses' arms in the $x$- and $y$-direction indicate the
width of the $\ell$ range and the measurement errors of the power,
respectively.}
\end{figure}

When we apply all the constraints mentioned in
Section~\ref{sec:parameters}, the only cosmological parameter that
remains free is actually the redshift of reionization $\zr$.  From the
$\chi^2$ analysis, we obtain a confidence interval of $18<\zr<25$ at a
level of 68\% and a range of $16<\zr<29$ at 95\% confidence.
Figure~\ref{fig:maxima_bestfit} compares the data themselves and the
model that provides the best fit.  This fit is clearly acceptable,
having a $\chi^2$ of 8.3 for ten degrees of freedom.

As explained above, we use various observational constraints on
cosmological parameters.  To illustrate the effect of errors in their
measurement on our results, we subsequently omit each of them in turn.
This means that we have a second free parameter, in addition to the
redshift of reionization.

\begin{figure}
\includegraphics[width=\linewidth]{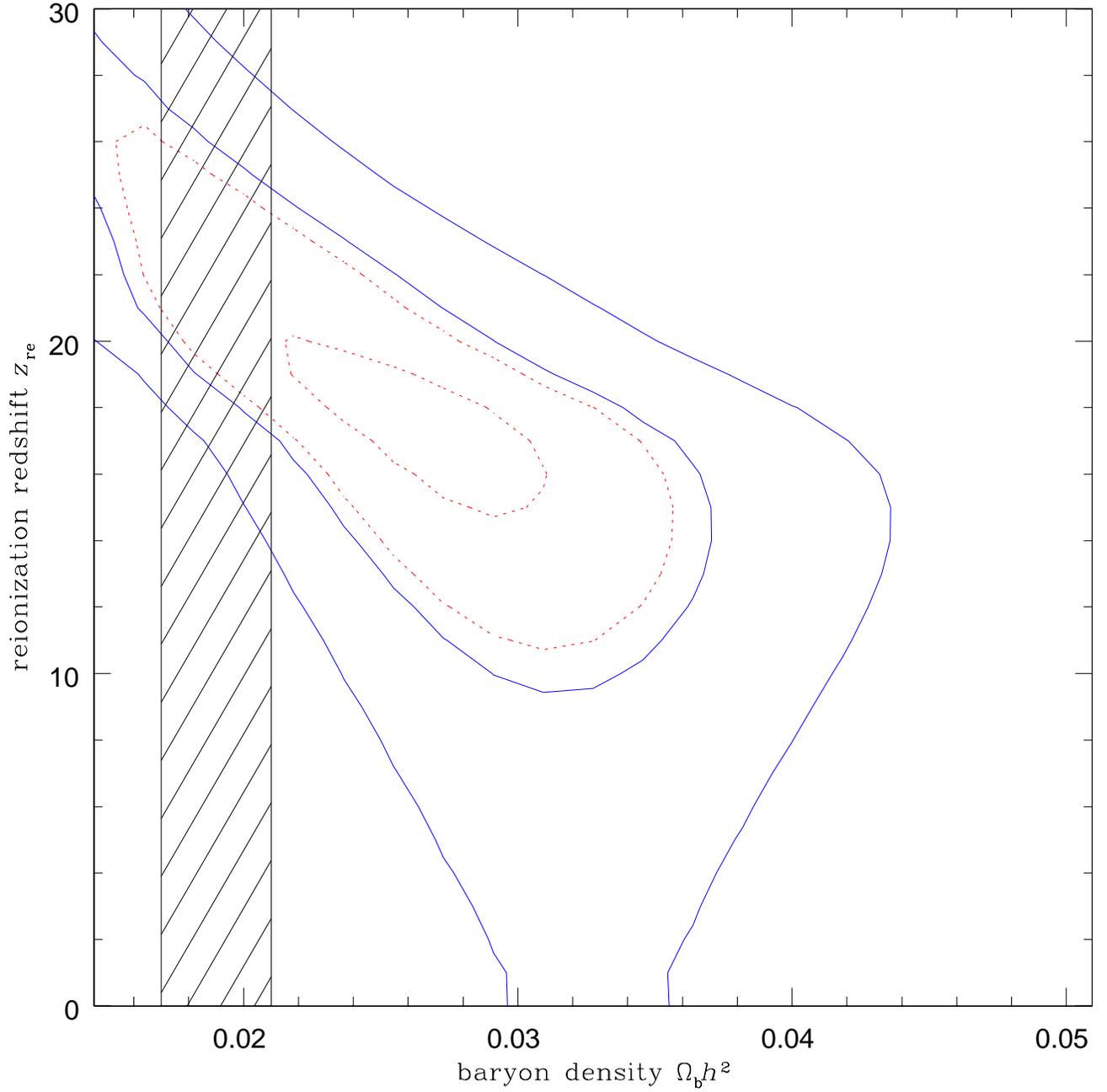}
\caption{
\label{fig:maxima_omegab}
Likelihood contours for varying baryon density $\Omega_\rb h^2$.  The
Hubble constant is fixed at $h=0.65$ and the
constraint~(\ref{eq:baryonnicfraction}) on the baryonic fraction
$f_\rb=$ is in effect.  The solid blue lines indicate the one- and
two-sigma error regions determined from the $\chi^2$ values.  The
dotted red lines show the same for the maximum likelihood analysis.
The shaded bar indicates the observational $1\sigma$ range of
$\Omega_\rb h^2$ obtained using BBN.  }
\end{figure}

\begin{figure}
\includegraphics[width=\linewidth]{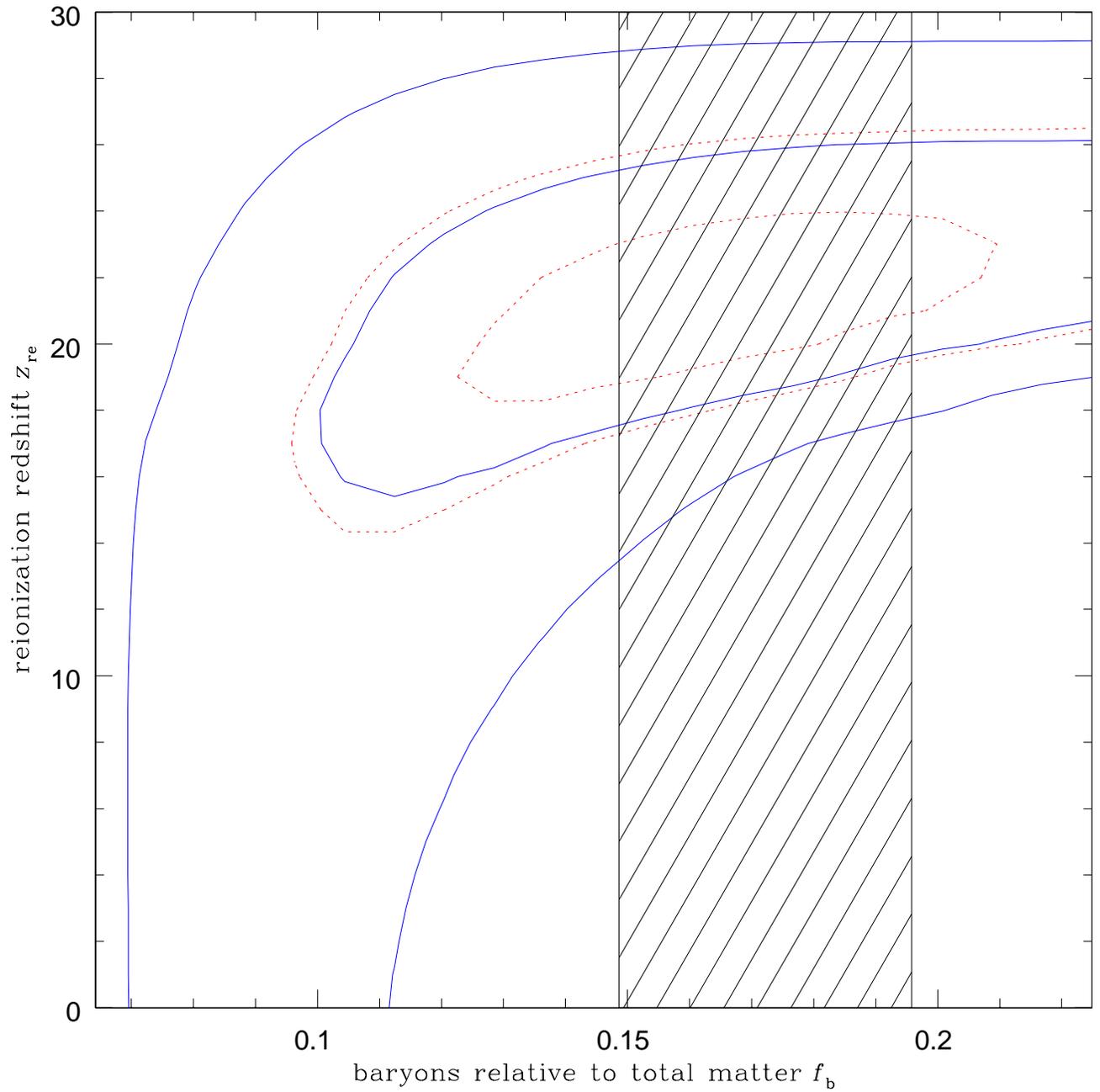}
\caption{
\label{fig:maxima_fractn}
Likelihood contours for varying baryonic fraction $f_\rb$.  The Hubble
constant is $h=0.65$ again, but now $\Omega_\rb h^2$ is fixed at
$0.019$.  Line styles are the same as in the previous Figure.  }
\end{figure}

\begin{figure}
\includegraphics[width=\linewidth]{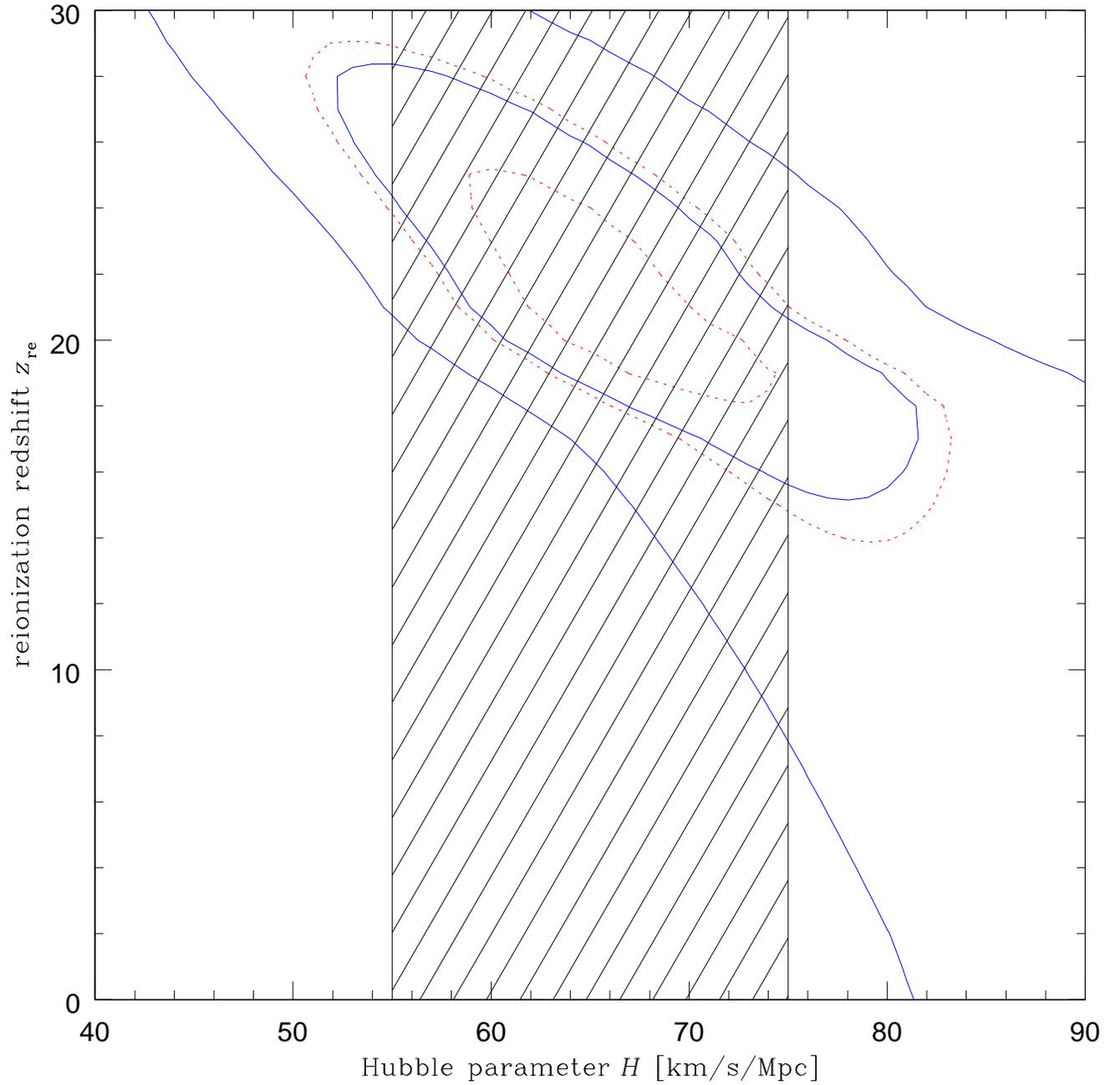}
\caption{
\label{fig:maxima_hubble}
Likelihood contours for varying Hubble constant.  Both $f_\rb$ and
$\Omega_\rb h^2$ are now constrained as explained in
Section~\ref{sec:parameters}.  Again, solid blue and dotted red lines
indicate the error regions for the $\chi^2$ analysis and the maximum
likelihood method, respectively.  }
\end{figure}

In order to illustrate the methodical point made in
Section~\ref{sec:method}, in Figures~\ref{fig:maxima_omegab},
{\ref{fig:maxima_fractn}, and {\ref{fig:maxima_hubble}}} we also show
the confidence regions obtained from maximum likelihood analysis.  It
is interesting to see how much smaller they are compared to the
regions obtained from the $\chi^2$ values.  We interpret this as an
indication that one should be cautious and use the larger regions, in
order to not rule out perfectly valid models.

Figure~\ref{fig:maxima_omegab} and Figure~\ref{fig:maxima_fractn}
display the constraints on the redshift of reionization obtained when
the baryon density $\Omega_\rb h^2$ and the baryon fraction $f_\rb$,
respectively, are allowed to vary.  Of course, this variation somewhat
broadens the allowed redshift range.  However, the change is not very
large, in either case the lower bounds on $\zr$ go down to $17$ and
$13$ for the 68\% and 95\% confidence level, respectively.

The situation changes somewhat when we let the Hubble constant vary
around the adopted value of $h=0.65$.  Figure~\ref{fig:maxima_hubble}
shows the results.  Despite the rather large uncertainty of $h$, we
can still constrain $\zr$: Unless $h>0.75$, which seems unlikely when
combining various methods of Hubble parameter estimation, we can
conclude that $8<\zr<32$ at the $2\sigma$-level, and at the
$1\sigma$-level we find that $15<\zr<27$.

\section{Discussion and Outlook}
\label{sec:discussion}

We have used the Maxima measurements of the angular power spectrum of
the CMB to constrain the redshift of reionization $\zr$.  We found
that astrophysically interesting bounds can be found when one
constrains the allowed range of cosmological parameters with other
observations, including Big Bang Nucleosynthesis and X-ray cluster
data.  Even when the possible uncertainties of these parameters were
taken into account, we still obtained a robust estimate $8<\zr<32$ at
the 95\% confidence level, unless $h>0.75$.  The implications of this
result for structure formation theory and the mass of the putative
warm dark matter particle will be discussed in a forthcoming paper.

One caveat is our omission of the possible contribution from tensor
modes: tensor modes (gravity waves) contribute extra power on
super-degree scales ($\ell<100$), so that when the total power (scalar
and tensor modes) is normalized at low $\ell$, the scalar modes become
lower and thus the high-$\ell$ power looks suppressed.  That effect is
similar to that of reionization, but fortunately this degeneracy can
be broken with future observations of the polarization angular power
spectrum.  For an electron scattering optical depth of $\tau=0.1$ --
corresponding to $\zr\approx13$ for standard cosmological parameters
-- the polarization signal should be easily detectable by the MAP and
Planck missions {\citep{bennett1995,bersanelli1996}}, and possibly
also the forthcoming new Maxima and Boomerang missions.

\section*{Acknowledgments}

We have benefited from discussions with Per Rex Christensen, Gus
Evrard, Pavel Naselsky, and Joe Silk.  This work was supported by
Danmarks Grundforskningsfond through its support for TAC.

\bibliography{bibliography}

\onecolumn

\setcounter{section}{0}

\end{document}